**Spontaneous Formation of Universes from Vacuum via Information-induced Holograms**
B. Roy Frieden (1) and Robert A. Gatenby (2)
((1) Optical Sciences, University of Arizona, Tucson, Arizona, USA, (2) Depts. of Radiology & Mathematical Oncology, Moffitt Cancer Center, Tampa, Florida, USA)

All spontaneous emergence of quantum particles from false vacuums can occur via usual energy-based Lagrangians; or, as we show, via a variational principle of minimum loss of Fisher information. By this principle *all material existence* in the multiverse, including its life forms, are physical manifestations of Fisher information. The information principle serially formed our universe, and all others, in the multiverse. The resulting expansionary (Big bang) eras of time t and/or space-time x_i, i = x,y,z,t (c=1) for the universes are found to obey probability densities p(t) and p(x_i) of usual exponential forms. The existence of the multiverse allows preservation of invariant values of the 26 physical constants via their relay from one universe to another by successive Lorentzian wormholes. At each relay the emerging constants are represented by the intensities of an input hologram. The information principle was previously used to derive nearly all textbook physics and much cell biology, e.g. the Hodgkin-Huxley (H-H) equations governing ions emerging into *biological* cells. The equations we derive governing p(t) and p(x_i) for universes coincide with the H-H equations governing ions entering these biological cells. Thus, the information concept holds over a vast range of scale sizes.

## 2. BACKGROUND

Note: By "particle" we mean the usual wave-like entity consisting of energy-mass.

We propose that all statistical, scientific (physical, chemical, biological, medical) effects are tangible expressions of maximized *information* [1],[2]; in particular of Fisher information[3],[4]. These may be viewed, alternatively, as accomplishing *minimization* of its *loss* [5],[6] after transmission of certain particles over a channel. Also, in cases where the emergent medium is *granular* the Fisher information goes over into a Kullback-Leibler divergence value [1] Eq. (A4) or (A6) (see Appendix A).

In turn, this minimized loss is shown to represent that in corresponding *Shannon* information values [5],[6] (see below Eq. (3)). How does "minimum loss of information" occur?

### 2.1 Criterion of minimum information loss

A general information channel consists, by definition, of a signal, e.g. a light beam which is transmitted over a medium (called a "channel"), e.g. "the air," from its source (say, a time-modulated light bulb) to a receiver (say, your eye). The ideal "information" sought at the receiver is in the true time-dependent intensity profile at the bulb. The transmission channel is, by definition, closed to 'outside' *signal* inputs, so that, en route to the receiver *no new information* (here, light) *may enter it*. Then what *can* enter must be, by definition, purely noise. This can only reduce the source level of information. Hence, any change in the information carried from source to receiver *must be a loss*. Minimizing this loss then amounts to achieving maximum gain of the information over the channel.

Moreover, an effect whose observations convey *maximum* Fisher information *I*, in particular, can be *most accurately measured*. This is by the Cramer-Rao [1],[2],[3] result

$$e^2 = I^{-1} \qquad (1)$$

for the minimum mean-squared error $e^2$ attainable in measuring a mean, or signal, value. Thus, the larger the information *I* is the smaller is the mean-squared error.

## 2.2 Thesis of J.A. Wheeler

This has further consequences: Man is an integral part of nature, and *any humanly-observed phenomenon is, at least, affected by man's chosen method of observation* (a well-confirmed thesis of J.A. Wheeler [7]). An oft-quoted example is the two possible types of output – either visible interference *fringes* on a receiving screen or visually observed photon *slit positions*—when conducting the famous optical double-slit interference experiment [8]. In fact all such Wheeler-type experiments could be observed by suitably automated measuring devices and, so, are *not* necessarily limited to human – or even to live - observation. Apes, dolphins or artificial devices - such as motion-sensing cameras – can fill this role as well.

## 2.3 Dual role of intelligence

Next, consider a different benefit of quantifying an unknown scientific effect by *maximizing* its level of Fisher *I* [1],[2]. By Eq. (1), the *data* from the effect *are also maximally accurate*. In summary, in a universe such as ours maximum intelligence is built into *both (1) the structure of its physics and (2) the ability of that structure to be known.* This also lends itself to conveying into our universe extremely accurate values of the 26 fundamental physical constants, as were needed for it to evolve properly from the Big bang onward (see below).

Also, in such a universe, creatures that can deduce properties (1) and (2), and use them for evolutionary advantage, are favored. An example of such use is the current development of "learning machines." Suggestions of combining a learning machine with the human brain are being considered for, potentially, 'engineering' a combined living-digital being of increased mental faculty.

## 2.4 Accuracy problem

Could nature per se, in even the seemingly most complex form of multiple universes (a "multiverse"), have emerged out of this Fisher information-based property? A clue is that Fisher information is *a local measure* of complexity (in time and space), so that maximizing it allows, *in turn*, each universe of a multiverse to arise. We will show below how this could have happened.

A related question is how our universe could have evolved with the required extreme accuracy of the 26 universal constants that enable life to exist in it. For example, Stephen Hawking has noted that:

"The laws of science, as we know them at present, contain many fundamental numbers, like the size of the electric charge of the electron and the ratio of the masses of the proton and the electron. The remarkable fact is that the values of these numbers seem to have been very finely adjusted to make possible the development of life."[9]

## 2.5 Sensitivity of the known physical laws to accuracy of their universal constants

Thus suppose, e.g., that the strong nuclear force coupling constant was 2% stronger than it is, while the other constants were left unchanged (discussed by physicist Paul Davies [10]).

This "2% discrepancy" would leave diprotons stable. But then hydrogen would fuse into them instead of into deuterium and helium. This would drastically alter the physics of the universe and, presumably, rule out the existence of life as we know it on Earth. There must, therefore, be a mechanism for preserving to extreme accuracy the values of the 26 fundamental physical constants. Our model proposes such a mechanism (see below).

### 3. SPONTANEOUS ORIGIN OF INFORMATION

Consider the 'perfect' vacuum that existed prior to the Big bang expansionary phase. *We regard the Big bang onset of our universe A to be merely a large-scale version of the usual "emergence from false vacuum" effect* [*11*].By the latter, even a seemingly 'perfect' vacuum allows particles, such as electron-positron (e-p) pairs, to spontaneously arise from it, provided they mutually annihilate an instant later (since their existence violates *conservation laws*). Thus the 'perfect' vacuum allows such particles to emerge, albeit temporarily. It is therefore called a 'false perfect' vacuum or, simply, 'false vacuum.'

The preceding is called a "metastable state" of the vacuum, and it occurs because it has a small but finite probability of occurring quantum mechanically. And by the law of large numbers [4] if an event has a finite probability of occurring it *will* eventually occur. Precisely when, depends on the time scale of the system. See below.

In fact, in the presence of two other, *pre-existing* electrons the emergent e-p pair particles *can exist indefinitely* (since their presence allows the above conservation laws to now be obeyed). We will find, below, a very different condition for not only the emergence of the e-p but their stable existence thereafter. This is the role played by information, in particular of the Fisher variety.

Returning to the problem of generating a universe, it will be seen that *before* the Big bang occurs *its scales of time and space are undefined*. Thus, the Big bang serves *to define* these scale sizes. This serves to resolve the dilemma that actuation of a Big bang might have required more elapsed time than has already passed in our universe (call this the "time dilemma"). In effect, no time passed at all until that Big bang occurred.

Thus, we require a Big bang that *defines* the scale of time. This is by requiring time values that *convey maximum information I* about their values. In general, requiring maximum Fisher *I* gives laws $p(t)$ that tend to be maximally concentrated or *collapsed about* their means. This also tends to minify the above "time dilemma." Specifically it gives rise (see below) to the requirement of a family of exponential laws $p(t)$ on the time. Such laws satisfy other cosmological requirements as well (see below). Hence it is reasonable that the Big bang occurred.

### 3.1 Why is the Big bang an expansionary (not contractive) process?

(The prescient and very readable Ref. [11] is used throughout this subsection.) A vacuum is ordinarily thought of as empty space, but modern particle physics regards false vacuum as a physical object, having gravitational energy density. Also, the gravitational force is repulsive so that *the vacuum exerts outward pressure*. As with any quantum object, it can be in a number of different states of the vacuum. The higher is the energy of the vacuum, the stronger is the repulsion. As with electron-positron emersion from imperfect vacuum, this kind of vacuum is unstable. It decays into a lower-energy vacuum, and the excess energy produces an expansion of space and time (analyzed here); it later produces a fireball of particles and radiation (*not* analyzed here). The existence of false vacua follows from particle physics and general relativity, distinct from considerations of inflation.

Thus, the theory of inflation assumes that at some early time in its history the universe occupied a high-energy false vacuum. Repulsive gravitational forces then caused exponential expansions of space and/or time (found below) of the universe. Depending on assumptions, in about 330 doubling times the universe grows by a factor of 10-100. Thus, regardless of its initial size, the universe very quickly becomes huge. But the expansion has an end: because the false vacuum is unstable it eventually decays, producing a fireball. Our analysis is of inflation per se, *and therefore ends just before the fireball event* (*Note*: In fact our information-based approach gives, as well, the familiar exponential probability law for the fireball process as well, but is not shown here for brevity.)

First let us try to find the probabilistic behavior of the expanding universe at a general time *t*, with the Big bang defined to have occurred at a time $t \approx 10^{-36}s$. This defines the expansionary time interval, or era [12],[13] of cosmological evolution as $(10^{-36}s, T)$ with $T \approx 10^{-32}s$. (For simplicity we replace $10^{-36}$ by *0*). The time-scale (and space-scale) were actually expanding during that interval.

**3.2 Possible limitations to the theory set by the cosmological principle (cp)**

By the translational invariance property of the Friedmann-Robertson-Walker *cosmological principle* (cp) [12],[13],[14], all space positions $(x, y, z)$ are *equivalent* (all see statistically the same universe; but also, see a violation of this below--the directionality of the cosmic microwave background CMB radiation). On this basis, the probability on space and time is not of (i) jointly a definite position and time $p(x, y, z, ct)$ (as usual with $c \equiv 1$) but, rather, of (ii) a *purely temporal* process $p(t)$ occurring equally at all space values $(x, y, z)$. We alternatively take both points of view (i), (ii) in this paper.

Hence, we first consider the probability law $p(t)$ on the emergence times of particles entering universe *A* from *B* at entrance gate *P*. (Note: This does not violate the premise that particles of mass intrinsic to *A* do not form until *after* the expansionary era; since *these* particles had already formed in, and were intrinsic to, *B*.) We postulate that at this primitive stage of evolution the probability law $p(t)$ represents, more generally, that on the time values of *any* well-defined sequence of observations *t* in *A*. In this sense, then, $p(t)$ is the probability law *on the random variable 'time' itself*. The time values *t* are the arrival times of the 26 physical constants from universe *B* (see below).

**3.3 Case at hand**

As an example of such observations, these could be those of the positions of the matter-energy particles from universe *B* as they attempt to enter universe *A* precisely at gate *P* (see set-out Eqs. (2) below). An analogous situation arises in cell biology when ions attempt to sequentially enter a cell through a (biological) gate in its plasma (outer) membrane [5],[6]. There the $p(t)$ represent those on the entrance times of the randomly entering ions. And the "emergence" into the cell is the event that the ion simply passes *through* the gate in the membrane.

Also, for convenience of notation, the beginning of the expansionary era, defined to be when the first particle from *B* enters *A*, is denoted as 0 instead of its approximately known value $10^{-36}s$ (see above). Then $p(t)$ represents the probability of the occurrence *time t* of some *(any)* definite event as observed in some sequence (see below for specific examples).

### 3.4 Exceptions to a non-directional nature of the cosmological principle

However, other approaches emphasize the *limited* validity [14a,b] of the invariance property of the cp. For these scenarios *the universe does have definite directional properties*, so that the *fully* joint probability $p(x_\alpha)$ on 4-position $x_\alpha = x, y, z, ct$ (with $c = 1$) on position and time has *physical meaning*. In these cases the probability is a usual *covariant* function of general relativity; showing valuable self-consistency. In fact we shall take, alternately, *both points of view*; first deriving the form of $p(t)$ and then showing how $p(x_\alpha)$ likewise follows. Both are based on principle (5) (below) of maximum Fisher information (also previously called principle "EPI").

The results will also be seen to satisfy both requirements of (a) accurate duplication of the 26 physical constants and (b) formation out of a hologram (located at the entrance to universe *A* set out in Eqs. (2) below. The hologram serves to effectively transmit the values of the 26 constants from universe *B* to *A*. More generally, each universe in the multiverse is taken to form the constants *from a predecessor universe*, as in Eqs. (2).

These are properties of the following model, which is proposed to underlie formation of our universe.

### 4. EMERGENCE OF OUR UNIVERSE *A* IN A MULTIVERSE

We view all existence, including that of our universe *A*, as being within an overall Guth multiverse [13] *G*. This allows universe *A* to have formed, out of imperfect vacuum, from some other universe *B* within multiverse *G*, as below. Only by the multiverse mechanism could it obtain (via *B*), all 26 required physical constants [15a] and the Higgs boson (or its field) with virtually perfect accuracy. This is by the following, recently published, wormhole-based model [15b].

### 4.1 Wormhole-based model

i. In a neighboring universe *B* to ours we consider a finite region which does not realize the minimum energy of a true vacuum. In Fig. 1, B is that universe.

ii. Classically the region can be stable despite its raised energy, and is thus called a 'false vacuum'. But quantum mechanically the probability that the false vacuum decays to the true vacuum by quantum tunneling is non-zero, so that this process *will* occur, sooner or later. Of further interest is that quantum tunneling can be realized in two different ways:

iii. (a) *not a case of interest*: The region tunnels in *the same* asymptotic region that the starting false vacuum region was; then, the region expands at the expense of the false vacuum region. However this is within universe *B* (the upper body in Fig. 1) so that no new universe arises.

(b) *the case of interest*. Next, as a quantum process, the region (upper universe *B*) probabilistically tunnels, via a connecting pathway shown, *into a different* asymptotic region (lower universe *A* in Fig. 1) from the one inhabited by the starting false vacuum region. Thus, after the tunneling expansion takes place, the region that underwent tunneling evolves by growing its own space-time, *beyond the space-time region of its origin*. This 'outside' region later expands into a new universe *A* (ours). Or, probability law $q(x, t)$ goes over into law $p(x_\alpha)$ as in set-out Eqs. (2). Such expansion is in the spirit of the growing "baby" universes proposed by L. Smolin (see below, and Discussion). The umbilical cord in Fig. 1 is of unknown length but, for simplicity, taken to be zero in set-out Eqs. (2).

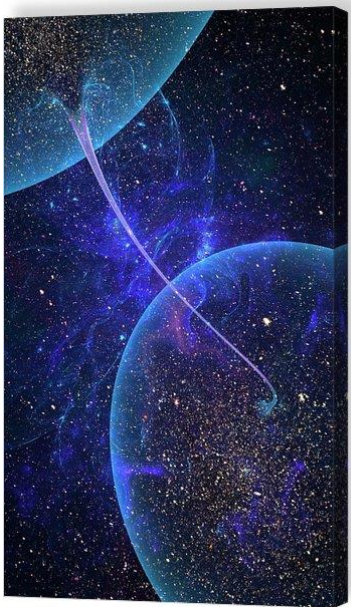

Fig. 1 Two universes: ours *A*(lower), and a neighbor *B* (upper), connected by an a Lorentzian 'umbilical pathway' Wormhole. The cord is of unknown length.

The entire region of growth is called a "Lorentzian wormhole." Further details are as follows. Regard the region in *B* that underwent tunneling to grow over some path (not shown, in the upper sphere of Fig. 1) to a point *P* on its 'bubble surface.' This wormhole generally acts to transport particles of mass-energy from the interior of *B* to its surface point *P* (by hypothesis during "past" times $t < 0$). From *P* these travel outward (downward) through the 'umbilical cord wormhole' shown in Fig. 1 connecting *B* to the newly forming universe *A* (ours) below. For simplicity we regard the surfaces of *B* and *A* to touch at the common point tangent point *P*. A cord of finite length has no essential role to play. See set-out Eqs. (2).

iv. Why do universes *A*,*B*,… form? The physical reason why universe *B*, which is assumed to already exist, branches off (gives "birth" to) a baby universe *A*, is that (as in the above emergence of an electron-positron pair from incomplete vacuum) it is a probabilistic, quantum effect which, since it has a finite probability of occurring, *does* occur (in effect, 'instantaneously' since time is not yet defined by an expansionary era in *A*). Extrapolating backward in time, in the same way a universe *C* gave birth to *B*, before that *D* gave birth to *C*, etc., starting with a first universe. Each *independently* underwent its own expansionary era, from initial conditions set by the 26 input constants, and independently evolved. By their independence, the Fisher information in the total sequence of universes is the sum of the information values. Also, by the principle of maximum Fisher information [1], that sequence must exist physically.

**4.2 *Lorentzian* wormholes realized**

These are essentially the time-development of three-dimensional wormholes [20],[21]. Such space-times exhibit wormhole structures on their space-like extensions. In general the wormholes either (a) connect one universe to another universe or (b) connect one region to another in the same universe. Indeed, both properties (a) and (b) are essential to our model. Thus, our Guth universes may well be Lorentzian in nature (see also Appendix B).

The above wormhole is assumed to transport mass-energy *particles,* including fermions and bosons (e.g. the Higgs boson). We suppose these particles to move outward from surface point *P* on universe *B* to a new universe *A*, within a coordinate system of ever-expanding coordinate range at times $T \geq t \geq 0$. Here $T$ is the end of the expansionary era at the space point $P$ and beyond. As abpve, we regard the bubble surfaces of *B* and *A* to be effectively tangent at the point $P$ (umbilical cord of zero length). By the cp the position of point $P$, and therefore the radius of *B*, cannot be known quantitatively. However, knowledge of $P$ —a space position— is not necessary to our analysis of the probability on *the time $p(t)$ alone* during the expansionary phase. However, derivation of the full probability $p(x_\alpha)$ does assume knowledge (in principle) of position *P*. We alternatively consider this scenario later.

## 5. CREATION OF UNIVERSES FROM FALSE VACUUM: ANALOGY WITH BIOLOGICAL CELL GROWTH

Our aim in this section is to quantify the above process of successive creation of universes. We proposed that *this process is an outgrowth of creation from (false) vacuum*; as in the case of electron-positron pair creation from such vacuum. As noted there, each e-p pair creation event requires the presence of two other, existent electrons. These two electrons act as "catalysts" in promoting the reaction, as do the protein gates in the analogous *cell membrane* case mentioned above (and in more detail below). We will show that the space-time coordinates $x_\alpha$ of the new universe *A* so created would temporally grow exponentially, as required in a Guth *multi*verse [12]. Also proposed is that the physical constants characterizing past universes are propagated into universe *A* via an intensity hologram $q(t)$ within a neighboring universe *B*; and that the violent creation of *A* coarse-grains its space-time. The coarse graining allows an important simplification in the expression for Fisher *I* to result (see Appendix A).

### 5.1 Transport of the universal physical constants from Universe *B* to *A*

To quantify the above information-based model, the wormhole in *B* acts to transport quasi-particles of mass-energy density $q$ from the source within *B* to (say) a given, fixed output aperture *Q* on the surface of *B*, where $q$ emerges as a probability density function (pdf) $q = q(x_\alpha|K)$. Parameters $K \equiv c_1, \ldots, c_{26}, h$ are the values of all 26 (dimensionless) universal physical constants in *B*, plus fixed parameters $h$ that determine a Higgs boson field. The pdf flow $q(x_\alpha|K)$ thereby carries information determining the 26 constants and forming Higgs fermions through its dependence on vector $K$. The latter also represents initial conditions on defining the subsequent flow $q(x_\alpha|K)$.

How can flow $q$ physically express its $K$ vector after it emerges into our universe *A*? It first exits from aperture *Q* in universe *B* and then propagates through empty space into a new (our) universe *A* that spontaneously arises, and grows, out of this flow of particles. The vector $h$ permits the existence of mass in universe *A* via the Higgs particle. Universe *A* thereby obeys a new probability density law $p = p(x_\alpha)$ on space and time. Time *t* governs the expansion that has taken place up to that time. In summary

*B*➔*A*, **that is**

**mass-energy exits from *wormhole Universe B, via hologram-induced* [15c] *transitions from***

$$q\text{-----}to\text{---}\rightarrow\text{-----}to\text{--}\rightarrow\text{---}\rightarrow p \qquad (2)$$

*in our Universe A, defined over expansionary phase times* $T \geq t \geq 0$

(Also see Fig. 1). This is the process by which the 26 constants and $h$ are transported into our universe.

### 5.2 Specialization (temporarily) to time dependence alone

However, by the above *cosmological principle* cp, all space positions $(x,y,z)$ are equivalent. Thus, notation $x,y,z$ no longer has definite meaning, and can be deleted from notation $p(x_\alpha, K, t)$. The result is now *simply* $p(K,t) \equiv p(t)$, after also suppressing the important factor **K** for *brevity*. Thus, $p(t)$ means a pdf for time *t per se*, independent of (hypothetically) indistinguishable position values $(x,y,z)$. Interestingly, the corresponding problem in cell biology [5],[6] is likewise one of solely determining a probability $p(t)$ (see below).

Finally, although our universe *A* contains mass, this is only because of the existence here of the Higgs boson field **h**. Thus, the Higgs field was, likewise, passed on to us in the density hologram $q(x,y,z,K,t) \equiv q(t)$ for simplicity. In summary, to this stage the unknowns of the problem $p(t), q(t), K$ obey

$$p(x,y,z,t,K) \equiv p(t),$$
$$q(x,y,z,t,K) \equiv q(t) \qquad (3)$$
$$K \equiv c_1, \ldots, c_{26}, h$$

We often term these probability density functions (*pdfs*) as "flows," for convenience, although they obey normalization over time interval $(0,T)$ like any well-defined pdf. (Normalization is in fact used as an explicit constraint in forming the solution law $p(t)$ to the time problem; see (13) below) The time $t \equiv 0$ defines the start of the expansionary ("phase" or "state") of our universe *A*. What principle governs its time expansion $p(t)$? It is assumed that, at times $t \leq 0$ (while in universe *B*), space-time had a mass-density pdf of $q(t)$, that of the *reference hologram* later carrying knowledge of the 26 constants and Higgs field **h** in **K** and time *t into our universe A*.

How are times *t* physically defined? In this subsection we are assuming that universe *A* undergoes its expansion phase in time alone. As we saw at setout Eqs. (2), this expansion is formed by mass-energy particles traveling from the bubble exit of universe *B* into *A* (Fig. 1). But what physical values of *t* define $q(t)$?

**5.3 Times *t* identified as those at which the 26 physical constants are defined**

Recall, via Eqs. (2), that it is specifically the hologram $q(t)$ that contains information about the 26 physical constants). Hence the value of each such constant *may be carried by a distinct particle* traveling (as in Fig.1) from universe *B* to *A* during the expansionary era. This would define 26 values of *t.* (But see Appendix B for other strategies whereby, e.g., a smaller number of values *t* might have been used.) In any event, there was probably more than one expansionary event (Big bang) over the expansionary time era $(0, \tau)$.

The probability that the expansionary phase exists at some time value $0 \leq t \leq T$ of the expansionary era is $p(t)$. This thereby replaces $p(x,y,z,t,K)$, by the cp. Also, moments $< t^n >$ over all times of the expansionary era therefore obey

$$< t^n > = \int_0^T dt\, t^n\, p(t), \quad n=1,2,\ldots. \qquad (4)$$

**6. PRINCIPLE OF MAXIMUM FISHER INFORMATION**

The pdf $p(x,y,z,t,K)$ within *A* is postulated to arise out of a pdf $q(x,y,z,t,K) \equiv q(t)$ of mass-energy density that flows from its origin somewhere in *B* to a given, fixed output aperture *Q* on the surface of *B*. (See set-out Eqs. (2) above.) The wave *q* next exits *Q* (upper sphere in Fig. 1) and travels through perfect vacuum between universes B and A, via the 'umbilical' pathway shown, to just inside the entrance to A (lower sphere in Fig. 1). This is located an infinitesimal distance away from Q. It thereby delivers to A the 26 constants $K$ plus the Higgs particle or field.

### 6.1 Emergence from vacuum

Next, the wave just within the entrance to A *emerges there* "from vacuum" (see Eqs. (2)). However, vacuum space is far from empty: The emergence must be accompanied by sufficient catalyst particles to be made permanent (e.g. a pair of electrons in the above e-p pair emergence case). And it is conjectured that vacuum *energy* is the largest contributor to cosmological density [13],[14]. But from where do such particles arise?

An analogous emergence occurs (see below) in the *biological case of ion emergence into a detecting cell*. Wave *p* gains a maximum amount (or, to be precise, *loses* a *minimum* amount; see above) of Fisher information from that in wave *q.* The maximum is limited by local constraints, as is usual in variational problems (see below). These principles predict that each mass particle so acquired by universe A obeys the usual differential equation of quantum mechanics (depending upon conditions, either relativistic or non-relativistic, and with or without spin) [1].

### 6.2 Requirement of neighboring universe B

It will be seen that the resulting particle wave flow $p$ in universe A has a generally *exponential* time dependence. This is after emerging from the wormhole of B (as above, set out in Eqs. (2)), entering A and randomly filling each space position $(x, y, z)$ over all times $(t, t + dt) \geq 0$. (Recall this separation out of space from time events was mentioned, above, as due to the strong coarse graining of A that occurred during emergence of the mass-energy particle wave). The probability law so emerging is $p(t) \equiv p(t|x)$, i.e. the probability of event $(t, t + dt)$ at each *fixed* (but *un*specifiable by the cp) 3-space point $(x, y, z)$. As in the above case of an emergent electron-positron (e-p) pair, the emergence during interval $dt$ is only temporary: The pair combine and return to pure energy, *unless an equivalent number of real particles pre-exists it*. Hence we here assume that the wormhole link to B also allowed these particles to flow from B into A. The creation of universe A thereby depends upon the (*prior or joint*) *existence* of universe B (and the flow into A of some of B's particles).

Then, by the same token, the creation of B must have depended upon the (prior or joint) existence of another universe C; and this formed from a prior universe D; etc. Then a great many (possibly an infinity) of universes exist in the 'multiverse.'

This assumes that, as did A, each such universe emerged independently, out of its own Big bang from false vacuum.

A related question is how the newly formed law $p(t)$ in universe A gets to "know" the finely-tuned physical constants, mentioned above, that permit it to further evolve. As we discussed, these, likewise originate in the prior universe *B and flowed*, through a wormhole, *into A as the vector* $\equiv (c_1, ..., c_{26}, h)$ over times *t* (see Fig. 1). Also, since information *I* is maximized, by Eq. (1) the particles traveling from universe B into ours will be replicated with high accuracy, in fact sufficient to allow our physical laws (see above section "Accuracy problem") to be obeyed by very close to the universal constants that are observed nowadays in A.

The question of what formed the *first* universe containing such constants is therefore crucial; but, for brevity, is regarded as beyond the scope of this work.

**6.3 Analogous problem in cell biology**

It was shown [5],[6] by analogous math, that the exponential time dependence of *universe A* also defines the flow of ions (now in passage through the cell membrane of a *living biological cell*, as below). These ions transport, in this manner, vital environmental information to the cells (say, the location of a predator or prey). The resulting ion flows carry maximum information (about that location, e.g.), and thereby permit *optimally fast and accurate* communication among them. This promotes optimal survival rates (e.g., in deciding whether 'fight or flight' is indicated); and more generally the biological effect called *natural* selection [6] results. In this way, Darwinian evolution follows from the EPI principle.

Finally, as taken up below, the output of the principle of maximum Fisher information (in the presence of coarse graining) follows mathematical laws defining diffraction optics, with $q(x, y, z, t)$ acting as a reference "intensity" *hologram*. This is in accord with the thought [13],[14] that the details of function of our 3D universe follow the requirements of a 2D hologram. Such a hologram is presumed to also pass on the pre-existing values of the physical constants from universe *B*, via its wormhole, into our universe *A* (see above). Also as discussed there, each of these relayed values of the physical constants from *B* (including the Higgs boson field) is assumed to be received in *A*, as a form of *initial conditions* on the resulting expansionary era. These inputs to the problem are assumed to have sufficient accuracy (i.e. to provide sufficiently high enough Fisher information) to re-form (reproduce) the total physical theory of *B* as ours in *A*. In this manner the entire multiverse follows the same laws of physics. They are invariant, representing a kind of absolute multi-universal truth.

**7. FORMATION OF THE *MULTI*VERSE FROM FALSE VACUUM**

The premise of this analysis is that the matter-wave carries Fisher [1],[3],[4] *information* at a time *t* as transition

$$J \to I \qquad (5)$$

from, respectively, just inside the wormhole in universe *B* to just inside the entrance to our universe *A*. Both informations $I, J$ are about a particle *time* event *t* at any single, but (temporarily) unknowable *point* $(x, y, z)$. The left-hand side information $J$ theoretically exists just within universe *B*, i.e. over theoretical space-time values $|x| \geq 0, |y| \geq 0, z = 0$, at $t = 0$. And information $I$ exists throughout the right-hand space XYZ (of our universe *A*). These are over all *distinguishable* time values $t$ obeying $T \geq t \geq 0$ for some largest time value *T* of less than $10^{-30}$ sec. To reiterate: Time values *t* are assumed observable and distinguishable, but this is not true of space values $x, y, z$, as above. (However, the latter restriction may be lifted, as below.)

This picture of the Big Bang may be analogous to the way an electron-positron pair emerges randomly (likewise at $t = 0$) out of a local, empty vacuum - its so-called 'false vacuum.' (Note: this vacuum must be limited to temporary time existence, since it is known to be impossible [3] for the particle pair to so emerge without violating basic conservation laws.) However, as we discussed, if two electrons already exist nearby[3] the conservation laws are obeyed, and the existence can be long-lived. The two electrons then act effectively as *a catalyst* for emergence of the electron-positron pair. *Withou*t this catalyst the e-p pair mutually annihilates, and dissipates into pure energy (e.g., at low energies, gamma-ray photons).

### 7.1 Corresponding emergence of ions in biological cells

We are motivated by an analogous "emergence" event in biology [5],[6], in the Hodgkin-Huxley (H-H) scenario of an ion (say of $Ca^{++}$) particle passing through a neuron cell membrane, from the space outside the cell to the cytoplasm inside the neuron at a time $t$. Other such ions follow suit, likewise "emerging" within the cell at other times $t$, together forming a probability law $p(t)$ (as was described above for entrance time events of particles from universe $B$ into our universe $A$).

A catalyst is required at each such passage: a number of protein-based "gate" particles forming a circular opening in the membrane, thereby allowing the ion to pass through this membrane gate. In the present paper, the activity of the wormhole exit aperture of universe $B$ is modeled after this opening of the neuron membrane gate.

## 8. EMERGENCE OF CATALYST FOR FORMING THE MULTIVERSE

What was the catalyst for maximizing Fisher $I$ for universe $A$ during its formation from $B$ (and consequently for forming the entire multiverse)? This requires use of an effective catalyst, as in the above use of protein gate particles in the H-H cell membrane application. We next find such a catalyst for the case at hand of generating universe $A$.

### 8.1 Prior law $q(t)$ as a hologram

Then what 'catalyst' in the astronomical scenario would correspond to that of protein-based gate particles in the biological case? : The two universes $A$ and $B$ share the same fundamental physical constants, but develop independent of one another. Therefore the total Fisher information in both is their sum. Likewise, $B$ was formed by a prior universe $C$; etc. for previous universes. Therefore all contribute information *additively* to the total. Then, in the multiverse as a whole, what corresponds to "gate" particles is a potentially infinite regression of universes $A,B,C, …$, each contributing its Fisher information to a sum total value. By Eq. (1) the total information $I$ is to be maximized, and this is readily obeyed if, e.g., there is simply a "maximum" number of them (subject to appropriate constraints). On this basis, the catalyst for causing our universe $A$ to emerge from false vacuum is simply the existence of all prior universes $B,C,…$ And as we reasoned above, there is possibly an infinity of them.

However, another requirement of the Fisher principle exists, and that will be the *added term* Eq. (13) (see below) representing *prior knowledge* of the multiverse scenario. This serves to *constrain* the condition of maximum Fisher information in Eq. (13).

## 9. EPI PRINCIPLE

In the above Hodgkin-Huxley[5],[6] scenario, the ions *emerging into* a living cell from its plasma membrane obey a time flow $p(t)$ into the plasma obey a principle [1],[2]

$$I - J = minimum \qquad (6)$$

of "extreme physical information" or EPI. Such a flow of ions *delivers information of minimum loss* $I - J$ to the 'receiver' within the cell. (Note: "minimum loss" means in effect "maximum gain", as shown before) Such maximized information characterizes most textbook physical phenomena [1],[2]. Hence we use principle (6) here as well.

As will be seen, this predicts exponential increases in observed time values and/or space values depending upon whether the cp is assumed. But what kind of informations $I, J$ should be used?

There are two types possible: (1) Fisher information on the continuum; or (2) the Shannon information *limit* of the Fisher that occurs during coarse graining (derived in Appendix A). Note: In the limit of "coarse graining" space-time is sufficiently specified by increments of finite size. This may be due to, say, finite particles or time intervals being present. In the temporal case, the "coarseness" lies in the *intrinsic spacing* of neighboring *time* values being effected (observed). These might, e.g., be initially (at $t = 0$) spaced by 0.001 sec, but then by 1.001 sec, then by 10 sec, etc.

### 9.1 Effect of cosmological principle on the problem

To decide, let us consider what process is going on during the information transition from $J \to I$. The former, $J$, is the information level in space-time prior to the Big Bang. We now temporarily regard all four space-time values $\mathbf{x} \equiv \mathbf{x}_\alpha \equiv (x, y, z, t)$ as continuously knowable. Its corresponding Fisher information is, by definition,

$$I \equiv \int d\mathbf{x}\, \frac{|\nabla p|^2}{p}, \tag{7}$$

where $p \equiv p(\mathbf{x})$, $\nabla p \equiv \partial p / \partial x_i$, $\mathbf{x} \equiv \mathbf{x}_i \equiv (x, y, z, t)$, $d\mathbf{x} \equiv dx_i \equiv (dx, dy, dz, dt)$, $i = 1, ...,4$. Also $p \equiv \mathbf{q}^* \cdot \mathbf{q}$, $\mathbf{q} \equiv q(\mathbf{x})$, where $q(\mathbf{x})$ is a generally complex amplitude function, i.e. containing a definite phase function.

With *J* the information level in space-time prior to the Big bang, at *t* = 0, an instant later the Bang occurs, whereby chunks of mass-energy are *randomly* produced and likewise the time intervals between successive observation are random. As we discussed, by the cp these exist as well-defined observables *in time*, but not necessarily in space. However, if general relativity is to remain covariant *in principle,* we must regard the space coordinates as, likewise, knowable *in principle*.

Also, no well-defined phase function exists since the scenario is assumed coarse grained and strong decoherence enters in. In this limit the Fisher information goes over into Shannon information (see Appendix A). Also, as mentioned above, by the cp condition the four-dimensional $(x, y, z, t)$ information *I* of Eq. (7) effectively goes over into one-dimension, in time $t$ alone. Thus, under the cp model, where only time is a well-defined observable,

$$I \equiv \int dt\, {p'}^2/p, \quad p \equiv p(t), \quad p' \equiv \frac{dp}{dt}. \tag{8}$$

All integrals $dt$ go from $0$ $to$ $T$. We now return to the time-only expansion problem.

### 9.2 Friedman-Robertson Walker (FRW) solution

In the case we are considering, domination by the vacuum in the expansionary phase of the universe, the FRW model predicts that the expansion in time is exponential, as

$$a(t) \propto \exp\left(\sqrt{\frac{\Lambda}{3}} t\right), \tag{9}$$

where $\Lambda$ is the Einstein constant. However, $a(t)$ is not the probability law $p(t)$ we seek. Instead, $a(t)$ is the "cosmological scale factor". This scale factor is usually regarded as *a kind of* cosmological time measure, analogous to the probability law $p(t)$ we want. But in fact, our Fisher information-based solutions maximizing information (8) will likewise define an exponential *family* of solutions, whether or not the cp is assumed, and exponential form (9) turns out to be *its simplest member.*

### 9.3 Kullback-Leibler (K-L) divergence ('distance') measure

Continuing with dependence simply upon time *t*, the Kullback-Leibler (K-L) 'distance' measure $D_{KL}$ is

$$D_{KL} \equiv \int dt\, p(t) \ln\left[\frac{p(t)}{q(t)}\right], \tag{10}$$

(See Appendix A). The measure $D_{KL}$ is alternatively called the "Kullback-Leibler divergence", "relative entropy" or "weighted entropy" of $p(t)$, with $q(t)$ the K-L "reference-"- or "weight-" function. Most importantly for our purposes, it will be seen to form an *intensity hologram*.

In this coarse grained limit, the result (A4) of Appendix A follows for *I*. Comparing (A4) with definition (10) of the K-L measure gives

$$I = (2/\Delta x^2)\, D_{KL} \tag{11}$$

for a reference function $q(x) = p(x + \Delta x)$, $\Delta x \equiv (\Delta x, \Delta y, \Delta z, \Delta t)$.

Thus, by the 1-D *time* effect Eq. (10), the Fisher $I$ given by Eq. (11) is proportional to the Kullback-Leibler divergence $D_{KL}$ between the mass energy density function $p(t)$ at the entrance to universe *A* (see setout Eqs. (2)) and its shifted version $p(t + \Delta t)$ an instant later. Also the information $I$ is seen to decrease quadratically (i.e. fast) with increasing coarseness $\Delta x$ of the emergent space. This makes sense, since it states that (a) the *more violent* was the big bang the *less information $I$* it formed in the space (universe) that followed. As further confirmation (b), by Eq. (1) the decreased $I$ means larger mean-square error in locating the particle.

### 9.4 Temporary return to full space-time problem

This occurs, e.g., when regarding the lack of knowledge of the 4-dimensional *origin* of space-time as merely one of *temporal* ignorance (i.e., currently we simply do not know it) rather than one of physical impossibility. Then information $I$ in Eq. (7) is the Fisher information. In full space-time $x_\alpha \equiv x$ ranges over all four-space positions $x_{\alpha_n} \equiv x_n$, so that information $I$ has the form

$$I = (2/|\Delta x|) \sum_n p(x_n) \ln\left(\frac{p(x_n)}{q(x_n)}\right). \quad q(x_n) \equiv p(x_n + \Delta x). \tag{12}$$

(See Appendix A) A case where full space-time explicitly enters in is when viewing the cosmological microwave background radiation [22] (CMB) in our universe *A*. It is known to suffer directional anomaly, *violating* the cp that the universe looks "the same" in all directions.

### 9.5 Temporary return to *time-only* problem

We want to apply this in Eq. (6) to ascertain the time distortion $p(t)$ after the big bang, i.e. over all time values $T \geq t \geq 0$ to the end $T$ of the expansionary era. It turns out (see below) that a *one*-dimensional form of Eq. (6) (solely over the time *t*) gives as the answer $p(t)$ a generalized exponential form for the time distribution $p(t)$. So far as is known, $T \approx 10^{-36}$ sec.

We recall that use of the principle Eq. (6) requires one constraint of *knowledge of what's going on physically.* In the H-H neuron case [5],[6] it was knowledge of a definite mean time <t> over which entering ions are within the cell membrane (CM). That is, the constraint was

$$\int dt\, t\, p(t) = <t>. \qquad (13)$$

Such an elemental constraint governs the 'strength' of the effect (the bigger $<t>$ is the bigger is its effect upon the solution $p(t)$). Hence, we shall later use this same constraint (13) on the model Eq. (6) for Big bang expansion during the *emergence* of our universe *A.* However there's a better reason for using constraint (13).

### 9.6 Constraints and cause of emergence

Of course, in our Big bang scenario for universe *A* we cannot know what constraint to enforce (nothing was there at the time to tell us what was happening) So let the constraint be more *generally* represented, as $\int dt\, f(t) p(t) = <f(t)>$, with $f(t)$ arbitrary. However, in fact it largely won't matter to the derivation what form $f(t)$ actually has.

Now we have to prepare for the bang (actually, a sudden expansion of the space of *t*). What sets it off? Again, we can't know. However, there is seemingly a known precedent for it, in the preceding case of electron-positron pair creation from the vacuum. There, two nearby electrons set it off (they *catalyze* the reaction, like equivalent 'gate particles' in the neuron case [5],[6]). Their presence allowed a very brief ep pair creation to become, now, long-lived; since long-term conservation of momentum and energy values could now be satisfied. Likewise, something happens to catalyze the Bang. Here it is (by set-out Eqs. (2)) the sudden appearance of one or more mass-energy particles out of a wormhole from universe *B*, or equivalently, the passage to *A* of a nearby matter wave. And as we saw, the existence of *B* requires, in turn, the existence of *preceding* universes *C,D,E,…* Thus all are necessary. And each independently operates to form its successor out of emergence from false vacuum (see above). Then the total Fisher information contributing to formation of *A* is their sum total information, a sum growing with universe generation number. This is favorable, but the probability of one or more errors in the 26 passed-on physical constants *also* increases with generation number. Recall that all 26 had to be present with extreme accuracy in order to physically evolve as we did. Therefore, eventually a universe might be formed whose constants are so erroneous that it does not evolve 'correctly' or, even, at all; terminating this particular sequence of universes (see also, below, the discussion of an evolutionary model of the multiverse by L. Smolin), suggesting that there might be a finite, optimum number of universes in the multiverse.

### 9.7 Continuing the 1-D case of time observation alone

To keep the presentation *simple* we again specialize to seeking solely the time-dependence $p(t)$ of the expansion. Consider the serial emergence of mutually interacting pairs … *D→C* then *C→B* then *B→A* of universes. Each forms out of false vacuum, in some definite time sequence, with its preceding universe acting as a formative hologram (here in the universe transition *B→A* alone). A resulting mass flow $p(t)$ over time ensues from the aperture at the entrance to *A* (see set-out Eqs. (2)). The arrival of these gate particles sets off a Big bang, forming a new universe over increasing time *t* values until a termination time $T$. The Big bang initiates at a finite time, but to keep the notation simple we call it a time 0. This flow follows our usual H-H principle Eq. (6) of minimum loss of information, plus Lagrange constraints, as in past [1],[2],[5],[6] applications of the principle:

$$I - J = \text{minimum} \equiv \tag{14}$$

$$\int_0^T dt\, p(t) \ln\left[\frac{p(t)}{q(t)}\right] + \gamma_1[\int_0^T dt\, p(t) - 1] + \gamma_2[\int_0^T dt\, q(t) - 1] + \gamma_3[\int_0^T dt\, f(t)p(t) - <f(t)>].$$

  K-L to be minimized            Normalization           Normalization      Prior knowledge $<f(t)>$
(Note: each of these four titles describes the bracketed [ ] term directly above it.)

Lagrange constraint constants $\gamma_i, i = 1,2,3$ multiply corresponding constraint factors in brackets [ ]. All these must be zero at solution (thereby satisfying the constraints). Notice that the first right-hand integral is the cross-entropy between $p(t)$ and its reference hologram $q(t)$. Since this is to be minimized, and $\ln(1) = 0$, the result is that $p(t) \cong q(t)$, so that the hologram form $q(t)$ strongly affects the time evolution $p(t)$ of the universe $A$. This is also called a 'matched' filtering answer: attaining the requirement whenever $p(t) \approx q(t)$.

As found before [1],[2],[5],[6], Eq. (14) is thereby also minimized subject to prior knowledge $<f(t)>$ (its far right-hand integral). In general, quantity $<f(t)>$ expresses prior knowledge of some definite average number describing the physically formative effect. As discussed prior to Eq. (6), here *the effect is emergence of our universe A* after acquiring, from prior universe B during the Big bang, the 26 physical constants defining its physics. As seen in the next subsection, this implies a definite value $<f(t)>$ of the mean.

Assume that a universe $A$-wide gravitational field dominates formation of the mean $<f(t)>$ used in principle (14). The latter describes interaction with $f(t)$ of the particle waves $p(t)$ traveling from universe $B$ to $A$. Mathematically, principle (14) is one of minimum Kullback-Leibler information, which is consistent with classical statistical effects [1] such as the 2nd law of thermodynamics. Thus our approach here for forming a multiverse is ultimately one of classical statistical mechanics. Its further significance is as follows.

### 9.8 Significance of the quantum-mechanical mean value

Here the requirement of "emergence" is satisfied by an "emergent", i.e. well-defined, mean value $<f(t)>$ of $f(t)$. That is, a quantum-mechanical mean value represents an emergent (classical) reality.

On this basis, reality $p(t)$ would result from requiring it to cause an average mass-energy flow $<f>$ of some instantaneous function $f(t)$ of the time, or even to obey simply the *simple average flow* over time. (In fact it *is* the latter, in the analogous scenario of ion flow through the cell membrane of a neuron [5],[6].) The form for $p(t)$ will be a *general exponential*, regardless of what particular function $f(t)$ is (unless it is a logarithmic function). An exponential answer is thus a strong one.

### 9.9 Continuing the derivation

First, the Lagrangian $L$ is defined to be the sum of all terms in its integrand that generally depend upon $p$ or $q$. From (14) this is

$$L = p(t) \ln\left[\frac{p(t)}{q(t)}\right] + \gamma_1\, p(t) + \gamma_2\, q(t) + \gamma_3\, f(t)p(t). \tag{15}$$

This lacking any terms in rates $p' \equiv \frac{dp}{dt}$, its Euler-Lagrange solution $p(t)$ obeys simply

$$\frac{\partial L}{\partial p(t)} = 0. \tag{16}$$

So differentiating Eq. (15) gives

$$\frac{\partial L}{\partial p} = 1 + lnp - lnq + \gamma_1 + \gamma_3 f(t). \tag{17}$$

By Eq. (16) the Euler-Lagrange solution is then

$$1 + \ln p - lnq + \gamma_1 + \gamma_3 f(t) = 0. \tag{18}$$

Solving this,

$$p(t) = q(t)\exp[-1 - \gamma_1 - \gamma_3 f(t)], \quad 0 \leq t \leq T \tag{19}$$

the general exponential spoken of above.

### 9.10 Multiplicative factor $q(t)$ as a hologram

Statistically, the multiplicative function $q(t)$ in Eq. (19) acts as a bias function for $p(t)$. Thus, such a function defines a kind of hologram. Lacking phase properties, $q(t)$ is an intensity hologram. Recall that by the cp nothing is known about preferred particle positions in this prior *space*. Therefore the hologram could well express maximum uncertainty in *time* as well,

$$q(t) = const \equiv q_0(\mathbf{K}), \ 0 \leq t \leq T. \tag{20}$$

As we saw, this hologram $q(t)$, arising as it does out of the neighboring universe B, carries into the law $p(t)$ of our universe A presumably *stable and implementable values of the 26 fundamental physical constants* as well as the Higgs boson field $\mathbf{h}$ (see above Eqs. (2)). This is through the constants $c_1 \equiv$ , ..., $c_{26}$, $\mathbf{h}$ . Without this hologram many physical and biological processes in A, especially on the micro-level, could become ill-posed. They would be dysfunctionally unstable in the presence of even slight uncertainties in their inputs (also see above).

Eqs. (19) and (20) simply combine as

$$p(t) = q_0(\mathbf{K}) \exp[-1 - \gamma_1 - \gamma_3 f(t)], \quad 0 \leq t \leq T \tag{21}$$

Thus the temporal mass flow $p(t)$ during the expansion phase of the universe is a general exponential function of the time. Through the holographic reference function $q_0(\mathbf{K})$ it also carries values $\mathbf{K}$ of the universal physical constants into the newly forming universe A. Regarding the time dependence of $p(t)$: By sight, if either $f(t) = \pm t$ Eq. (21) becomes a simple exponential law. We previously arrived at this form for $f(t)$ on the basis of definite emergence of a well-defined, mean value $< f(t) >$ of $f(t)$.

Exponential laws are known to be required in modern cosmology in order to satisfy the flatness and horizon effects required of the expansion. There is then physical significance to the case $f(t) = t$ and $\gamma_3 = -\sqrt{\frac{\Lambda}{3}}$, since in its $t$−dependence the directly exponential solution

$$p(t) \propto exp\left((\sqrt{(\Lambda/3)}t\right), \quad 0 \leq t \leq T \tag{22}$$

results. This confirms the "model exponential law" Eq. (9). The proportionality constant is such that $p(t)$ obeys normalization over time interval $0 \leq t \leq T$ defining the expansionary phase. A justification for the particular choice of constant $\gamma_3 = -\sqrt{\Lambda/3}$ follows.

### 9.11 Comparison with the standard "cosmological scale factor"

Eq. (22) for the probability density $p(t)$ has, in fact, the exponential time-dependence

$$a(t) \propto exp\ (bt),\quad 0 \leq t \leq T,\quad b = \sqrt{\Lambda/3} \tag{23}$$

of Eq. (9), the *cosmological scale factor* $a(t)$ of the Robertson-Walker solution for emergence from pure vacuum. Parameter $\Lambda$ is the Einstein constant. The scale factor is usually regarded as a kind of cosmological time measure. Result (22) independently confirms the identification in assessing the expansion of time *per se*. See discussion in introductory section (3.2).

As we found, effect (22) ultimately follows as a maximization of the Fisher information from before to after an aperture opens in universe *B*. This gave rise to universe *A*. Thus, as in the case of "spontaneous emergence" from vacuum of electron-positron pairs (see above), universe *A* arises as a "spontaneous emergence" from vacuum, the "vacuum" manifest as universe *B.*

### 9.12 Holographic properties of $q(t)$

The above development treats the emergence of the universe as an information-based process, whereby a source message $q(t)$ of low information content (the flat law (20)) is transmitted through vacuum space into an output space which will form our universe *A*. Aside from carrying the 26 universal constants into universe *A*, *in being flat* the law $q(t)$ is ideal as an initial version of $p(t)$ for our universe. Being so unbiased allows $p(t)$ maximum freedom of fluctuation, but realistically, subject to the extremum constraints enforced by fixed values of the 26 universal constants. In this sense, then, $q(t)$ is an ideal hologram for development of our universe *A*, forcing it toward being a stabilized version (property of a "*matched* filter") of the previous universe *B*. Randomly destabilizing effects are: the cosmic microwave background radiation level of 2.72548 K, which seems small enough to not destabilize further evolution of *A*; gravity waves (whose presence and largest possible magnitudes are, unfortunately, largely unknown); and explosive/implosive stellar events such as supernovae like SN1006 (reputably the brightest observed stellar event ever recorded). With the additional proviso that the Einstein constant $\Lambda$ remain constant, universe *A* seems potentially stable (at least, in its baryonic mass). However, little can be said at present about the stability of its dark energy and dark matter. Nevertheless, the existence of *multiple* universes of the Guth-Linde [12] type mitigates the extinction of a *lone* universe such as *A* -- perhaps such extinction would be as unimportant as the death of a single cell in a mammal. Also, while its baryonic mass might disappear the other two mass types might persist.

Hence, the required exponential solution for $p(t)$ is the outgrowth of a flow of matter-energy particles from universe *B* to ours *A*. These travel through Lorentzian wormholes (see above, and Appendix B). These are the counterparts of *microtubules* in biological applications [5],[6] of principle (6). This is to the scenario of ions entering a given cell in order (principally) to deliver environmental information, affecting survival, to the genes in its nucleus.

### 9.13 Generalization to pdf on *space-time* expansion after the Big bang

We now revert to the more generally covariant case where both absolute position and time in space-time are *knowable*. In this case Eq. (12) (or Eq. (A4)) gives four dimensions of information

$$I = (2/\Delta x) \sum_n p(x_n) \ln\left[\frac{p(x_n)}{p(x_n + \Delta x)}\right] \rightarrow (2/|\Delta x|^2) \int dx_\alpha p(x_\alpha) \ln\left[\frac{p(x_\alpha)}{p(x_\alpha + \Delta x_\alpha)}\right] \quad (24)$$

on the continuum. But in fact there are many special cases wherein an absolute origin in space *can be constructed*, so that, therefore, a probability law $p(x)$ governing space *and time* expansion after the Big bang has at least limited physical validity. Can the preceding derivation be generalized to include space variables $(x, y, z)$ as well? Yes, quite easily. See the Appendix.

## 10. SYNOPSIS

The 26 universal constants plus Higgs constants are relayed, via the hologram, and encoded as their respective expansionary times $t$, and space-time arrival measurements $x$. These time and space values obey exponential probability laws $p(t)$ and $p(x)$, agreeing with the usual exponential expansion of time $a(t) \propto exp((\sqrt{\Lambda/3}\, t)$, but also covariantly predicting an expansion of space-time

$$p(x_\alpha) \propto exp\left(\left(\sqrt{\frac{\Lambda_x}{3}}(x_\alpha - x_{\alpha 0})\right)\right) \quad (25)$$

(cf. Eqs. (22), (23) for time alone). These space-times are quantum outgrowths, on a massive scale, of vacuum energy as in electron-positron formation from vacuum. Location $x_{\alpha 0}$ of the center of our universe *A* and the constant $\Lambda_x$ are currently unknown; and $\Lambda$ is the Einstein constant of general relativity.

## 11. RELATED CONCEPTS

### 11.1 On the confluence of the coarse-grained nature of quantum gravity and Fisher information

H. Matsueda [16] has shown that the Einstein energy-momentum tensor -- governing the theory of classical general relativity – can actually *derive directly from the Fisher information* defined for microscopic statistical data that are coarse-grained (as here). But this is for (equivalently) a Gaussian case of $f(t) \propto t^2$ rather than the linear case in Eq. (22). Nevertheless, this suggests that *gravitation*, like Fisher information, is fundamental to *learning* maximally, and takes part in *forming* most (all other?) physical effects through a principle of minimum average Einstein gravitational curvature. In fact, it is known that average gravitational curvature in our universe is close to *zero*. Hence learning physical laws from observing gravitational curvature would seem to be an ill-posed problem. However, gravitational matter waves have recently been observed [17],[18],[19] and deducing known physical laws from the equivalent levels of Fisher information have already been accomplished [1], so that the gravitational approach to such deduction seems plausible.

### 11.2 Alternative approach to astronomical emergence
In a different approach [23], the Einstein–Cartan–Kibble–Sciama (ECKS) theory of gravity was used to explain why our Universe appears spatially flat, homogeneous and isotropic.

### 11.3 Recent approach to universal gravitation based on thermodynamic considerations

Here we showed how a multiverse of quantum-gravitational masses could form out of an entropy-like principle (14) of minimization of loss of Fisher (or Shannon) information. It was recently shown [24] that, analogously, use of the (non-exponential) Tsallis entropy in place of the exponential Boltzmann variety likewise allows gravitation *per se* to emerge as an entropic force.

## 12 DISCUSSION

### 12.1 Past derivations of physical laws using principle of minimum loss of Fisher information

Much textbook physics and biology was previously derived [1],[2] using principle (6) of minimum loss of Fisher information. But, were these merely mathematical 'tricks' for producing what the standard, *energy* based Lagrangian approach already does? Here we show that principle (6) derives, as well, the foundation law Eq. (25) for the expansion of space-time -- as well as a mechanism for preserving accurate values of the 26 fundamental physical constants throughout the multiverse. This seems to indicate that the information thesis gives a universal, alternative (to energy) meaning to physical existence in a multiverse: It favors the existence of systems (cosmological, biological) that maximally form, measure, transmit and express information, specifically Fisher information. In contrast with the phenomenological point of view of the usual least action approach, this information approach is also epistemological. It describes the unknown effect by its ability to convey knowledge - specifically maximum information - about its nature; rather than its ability to convey energy. (See Section (2.3) on "Dual role of intelligence.") Also, by Eq.(1), in maximizing information I it gives rise to maximally accurate measurements. This favors the existence of systems (biological creatures, systems of logic,…) that depend for their continued existence and evoluupon, as well, accurate inputs of information.

### 12.2 L. Smolin's 'biological' model for producing accurate values of the physical constants

We continue our biologically motivated, information-based approach for modeling the growth of a universe $A$ with correct (by definition) values of the 26 (plus Higgs) constants. Let these be passed on, basically unchanged, from corresponding "genes" (physical constants) of a "mother" universe $B$ to the "daughter" universe $A$. The mother's "umbilical cord" is, physically, the emergent Lorentzian wormhole as set out in Eqs. (2) and seen in Fig. 1, and is mathematically represented by the hologram $q(x)$ with its correct values for all fundamental constants of nature.

But first, a basic fact of thermodynamics: A universe eventually cools so much that it "dies" a heat death, i.e. loses the ability to drive "heat-driven" processes. The universe thereby reaches its state of *absolute maximum entropy*: all such processes cease. It is stone cold dead. This suggests a biological model for creation and evolution of a universe, as follows.

In a fascinating book [25], the physicist L. Smolin avoids the formal need for a hologram $q(x)$. He proposes that, instead, the problem of producing sufficiently correct 26 physical constants + Higgs constants can be effected by the following process of "natural selection."

Universes evolve by natural law out of the collapse of what are originally black holes. In turn, these universes "give birth" to offspring black holes (possibly via the "vacuum" emergence phenomena described here). Each, in turn, can become a universe, depending on the accuracy of its 26+Higgs constants, and the more accurate these are the higher is its "fitness." This is its probability of surviving long enough (by avoiding "heat death", see above) to, in turn, give "birth" to black holes; which themselves can (as in the preceding) become universes.

The closer are the values of the 26 constants in an offspring universe to the presumably ideal ones in our universe *A* the higher would be its probability of surviving. And so it would survive long enough to be on track for evolving into one or more universes, such as ours, that harbor life.

In this manner the collapse of a black hole could lead to the creation of a new universe. This daughter universe would have fundamental constants similar to that of the parent, though with some errors due to "mutations." And therefore a universe with sufficiently large parameter errors will reach "heat death" before being able to reproduce. Conversely, those parameters that postpone heat death the longest, in turn, survive the longest. As a consequence, certain universal parameters become more numerous, or likely, than others. These are, by definition, the 'correct' ones. Thus, this process ultimately accomplishes the "correct" 26 + Higgs parameters *out of a process of Darwinian natural selection*. This would seem to solve the problem of accuracy for the constants without, in particular, the need for a hologram. Instead, purely a process of 'natural selection' does the trick.

But, ultimately, it is fruitful to inquire what the origin is, *in the first place*, of the principle of *natural selection*. We previously found that the principle arises out of a principle of minimum loss of Fisher information [5],[6]. This is, in fact, also *the very principle assumed in this paper* (see Eqs. (6) and (14)) to facilitate *creation* of our universe *A* (subject to the existence of the hologram *q(x)*). So, both Smolin's route and ours follow from ultimately the same principle, that of minimum loss of (in fact, *maximized*) Fisher information. Through the temporal creation sequence of universes ...→ $C \to B \to A$ assumed in this paper, the entire multiverse runs on the information. This is, specifically, a "serial" creation of the multiverse.

However, universes $A, B, C, ...$ might instead have all arisen "in parallel", i.e. *simultaneously:* as one Grand Emergence event (... $and\ C\ and\ B\ and\ A$) from a grand imperfect vacuum. In principle, Smolin's approach could allow this (a case of total creation in a single generation), and *ours* couldn't: it is serial, as (... $C\ then\ B\ then\ A$). However, either process would have close to vanishing probability of producing such a simultaneous result, so any such comparison of the probabilities would seem inconclusive.

**APPENDIX A: RELATION OF FISHER I TO KULLBACK-LEIBLER K-L DIVERGENCE AND SHANNON INFORMATION**

We show results for, first, the general, covariant case 1 of space and time as observables. Then we take the cp case 2, where only *time t* is a well-defined observable.

**Case 1: generally covariant observation**

The continuous Fisher information measure $I$ is defined in Eq. (7) with $\nabla$ the four-dimensional grad operator. Measure (7) is most conveniently re-expressed in a discrete approximation

$$I = |\Delta x| \sum_n [p(x_n + \Delta x) - p(x_n)]^2 / p(x_n) \tag{A1}$$

with each $x_n = x_{n\alpha}, \alpha = 1, ...,4$. Also $|\Delta x|$ is the modulus of the tiny, fixed four-dimensional increment $\Delta x_\alpha$ characterizing the space. As a consequence of the smallness of $|\Delta x|$

$$\frac{p(x_n + \Delta x)}{p(x_n)} \cong 1, \text{ so that } \frac{p(x_n + \Delta x)}{p(x_n)} - 1 \equiv \varepsilon \tag{A2}$$

is small, so to second order, $\ln(1 + \varepsilon) = \varepsilon - \varepsilon^2/2$ or, equivalently,

$$\varepsilon^2 = 2[\varepsilon - \ln(1 + \varepsilon)]. \tag{A3}$$

Then by Eqs. (A2) and (A3), Eq. (A1) becomes

$$I = (2/|\Delta x|) \sum_n p(x_n) ln(\frac{p(x_n)}{p(x_n+\Delta x)}) \rightarrow (2/|\Delta x|^2) \int_0^T dx\, p(x) ln\left(\frac{p(x)}{q(x)}\right), \quad q(x) \equiv p(x+\Delta x) \quad (A4)$$

(in the continuous limit), plus two mutually *cancelling* normalization sums +1 and -1. The integral is the Kullback-Leibler (K-L) limiting form of the information *I*. (See also Ref. [1], pgs. 36-38 for details.) The right-hand side, explicitly representing Fisher information *I,* also represents the loss of *Shannon* information (SI) between space-time events $x$ and their displaced values $x + \Delta x$ (respective positions *before and after emergence from universe B*).

It is emphasized that, in (A4), no specific trajectories $x(t)$ are presumed. This is through the usual admonition that it is erroneous to assume, e.g., a spherically expanding wave of particles $r^2 = r^2(t) = (x - x_0)^2 + (y - y_0)^2 + (z - z_0)^2$ to exist. But this is actually beside the point. Each sum or integral in (A4) is a statistical expectation, so each foursome $(x, y, z, t)$ is just a random number *independently* substituted into distribution laws $p(x)$ and $q(x)$ in (A4).

**Case 2: limitation to time observation by the cosmological principle**

The cosmological principle states that the universe tends to look "the same" from all positions in space. Hence, the position coordinates of space-time events are not knowable. Effectively only time is. Then Eq. (A1) is, equivalently, replaced by $|\Delta x| \rightarrow \Delta t$, $x_n \rightarrow t$, giving information in time

$$I = \Delta t \sum_n [p(t_n + \Delta t) - p(t_n)]^2 / p(t_n). \quad (A5)$$

Next, steps (A2) and (A3) are taken, with vector coordinates replaced by the scalar coordinate of time *t.* Then result (A4) goes over into

$$I = (2/\Delta t) \sum_n p(t_n) ln(\frac{p(t_n)}{p(t_n+\Delta t)}) \rightarrow (2/\Delta t^2) \int_0^T dt\, p(t) ln\left(\frac{p(t)}{q(t)}\right), \quad q(t) \equiv p(t + \Delta t) \quad (A6)$$

The derivation of $p(t)$ and its ramifications continue at Eqs. (13)-(22).

**APPENDIX B: SYNOPSIS OF KEY REFERENCE**

There follows a synopsis of the key paper [20] by Li, Li-Xin. Note: *All references cited in square brackets [ ] in this Appendix are as originally in the Li, Li-Xin paper.*

Applications of wormholes in cosmology have also been investigated [9], [10]. In this paper the Lorentzian wormholes are essentially the time-development of *three*-dimensional wormholes [12],[13]. The wormhole space-time is constructed as a usual (classical) Friedmann-Robertson-Walker (FRW) universe. The FRW universe is open, with negative spatial curvature and can be extended with spatial hyperbolic hypersurfaces (H3). Each such H3 can be embedded in a four-dimensional Minkowski space-time through $t^2 - x^2 - y^2 - z^2 = a^2$, where (*x, y, z,t*) are the Cartesian coordinates in the Minkowski space-time and *a* is a constant. The metric of the Minkowski space-time is $ds^2 = -dt^2 + dx^2 + dy^2 + dz^2$ (*c = 1* as before). The open FRW universe has negative spatial curvature and can be extended with spatial hyperbolic hypersurfaces (H3). The resulting space-time consists of two open universes connected by a Lorentzian wormhole (our case of interest).

It is found that these wormholes could exist within a finite period of time *without* violating the weak energy condition [13],[14]. In this reference, the wormhole space-time is found to consist of two-dimensional space-like cross-sections with a topology of Tg (g ≥ 2), where Tg denotes a torus with g openings. For example, see Fig. 2, illustrating a case g=3 openings. *Thus, the author shows that the wormhole space-time represents two open universes connected by a Lorentzian wormhole* (our model choice). It has the following features: (1) It exactly satisfies the Einstein equations; (2) The weak energy condition is satisfied everywhere; (3) It has no event horizons, and thus *accepts all mass-like objects for shipment* from universe $B \rightarrow A$; (4) It has a topology of R2 × Tg (g ≥ 2), meaning it has g openings in its cross section (see case g=3 in Fig. 2). Also, its toral cross section narrows hyperbolically to a minimum at halfway between the two connected universes *A,B*; a typical such cross section case $g = 3$ is shown in Fig. 2. The minimum opening is not zero, so it accepts all candidate particles for shipment.

The value of g in the wormhole topology R2 × Tg is of *arbitrary* value (aside from obeying the constraint g ≥ 2). This allows a range of possible strategies for sending the 26 constants across the channel $B \rightarrow A$. For example a choice g =26 would allow *the 26 constants* to be sent, one to a chamber, *at a single time* across the channel. A potential benefit is that universe *A* thereby gets a quicker 'start' on its evolution than if (as in the above) the constants are sent one at a time. Many such transitions would seem more susceptible to error and, consequently, to suffer more error in the transmitted constants. Or, if g=13 is used, half the required number of constants could be sent, at each of 2 times. In fact there should be an *optimum* value for g, based upon requiring maximum total information in *A* over all received values of the 26 constants.

**Biological correspondence**

This cosmological Lorentzian wormhole case holds over colossal scales of distance. But it also has an analogous correspondence, on the microscale, in *biology to a (hollow) microtubule [5],[6]* with variable diameter along its length. The microtubule ships ions (such as the ubiquitous Na[++]) from one end, located at the surface membrane of a "cell B" (here *universe B*) to the other end, where it enters a given cell (universe *A*) through *its* membrane.

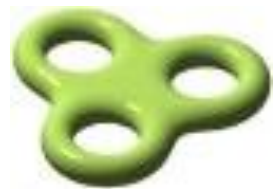

Triple torus g=3

Fig. 2 A typical cross sectional case *g=3* in a Lorentzian wormhole (Wikipedia).

Universe *B* thus provides a source, or "nutritive", environment (called a "substrate" in cell biology) for universe *A*. The "nutrients" so shipped are the 26 physical constants, received by *A* with maximum information at discrete times $t_1,…, t_{26}$ (or less, as discussed above). The utilization by *A* of each such constant thereby defines *a step in the evolution of A*. Also, the basic 26 physical constants of the astronomical application might have a cell-biological counterpart in, e.g., the 20 fundamental amino acids found in proteins.

A process expressing maximum Fisher information tends to form a computationally stable system (as demonstrated throughout [1] in derivations of standard textbook physics). Hence universe *A* tends to likewise evolve in steps comprising a well-defined, stable process. The information concept evidently holds over a huge range of size scales.